\documentstyle[12pt, fleqn]{article}
\begin{document}
\parskip=12pt
\baselineskip=20pt

\vspace{5cm}
{\begin{center}
{\Large{Spin-Gauge Theory of Gravity with Higgs-field Mechanism}} \\
\end{center}
\vspace{5cm}
\begin{center}
{{\sc H. Dehnen and E. Hitzer} \\
{\it Physics Department\\
University of Konstanz \\
Box 5560\\
D-7750 Konstanz}}
\end{center}
\newpage

 \section*{Abstract}
We propose a Lorentz-covariant Yang-Mills spin-gauge theory, where the 
function valued Dirac matrices play the role of a non-scalar Higgs-field. 
As symmetry group we choose $SU(2) \times U(1)$. After symmetry breaking 
a non-scalar Lorentz-covariant  Higgs-field gravity appears, which can 
be interpreted within a classical limit as Einstein's metrical theory of 
gravity, where we restrict ourselves in a first step to its linearized 
version. \newpage

\section*{I. Introduction}
Within the solar system and the binary pulsar PSR 1913 + 16 the
 classical gravitational interaction is described very well by
 Einstein's general relativity. However, this theory -
 simultaneously the oldest non-abelian gauge-theory with the
 Poincar group as gauge group - is not quantizable until now. On
 the other hand all the other fundamental interactions and their
 unifications are described successfully by quantizable Lorentz-covariant gauge
 theories with unitary gauge groups. Therefore the suspicion
 exists, that Einstein's theory represents only a classical
 macroscopic description of gravity and that the fundamental
 microscopic gravitational interaction between elementary
 particles is also described by a unitary gauge group on the Minkowski space-time in such a
 way, that Einstein's theory of macroscopic gravity is reached as an effective theory within a certain classical
 limit similarly as in the strong interaction the nuclear forces follow from the quantum chromodynamics.\footnote{
For this general intention see also Stumpf, 1988.
} In this way the problem of quantization of gravity and its
 unification with the other interactions would be solvable. 

In this connection the statement is of interest (Dehnen, Frommert and Ghaboussi, 1990), that the scalar Higgs-field of the elementary particle
 physics the basis of which are of course unitary transformation groups,
 mediates a Lorentz-invariant attractive gravitational interaction
 between those elementary particles which become massive by the
 spontaneous symmetry breaking, i.e. the Higgs-field has its
 source only in the mass and acts back only on the mass of the
 particles. The equivalence of inertial and gravitational mass is
 fulfilled automatically within this Higgs-field gravity. But if
 the strength of this gravity shall be of the order of the
 Newtonian one, the mass of the gauge-bosons will be of
 the order of the Planck-mass. 

For the last reason the standard Higgs-gravity, e.g. within the electroweak
 interaction (see Dehnen, Frommert, 1991), has presumably nothing to do with usual gravity. However, here
 the question arises, whether Einstein's tensorial gravity may be
 a consequence of a more sophisticated Higgs-field, which is
 especially not a scalar one. 

For this we extend back to a Yang-Mills $SU(2) \times U(1)$ 
spin-gauge theory of gravity on the Minkowski space-time of special
 relativity proposed by Dehnen et al. (Dehnen, Ghaboussi, 1985 and  1986 ; see also Chisholm, Farwell, 1989).  In this theory, where a subgroup of the
 unitary transformations of Dirac's $\gamma $-matrices  between their different representations (internal spin group; 
 see also Drechsler, 1988 and Bade, Jehle, 1953;
 cf. also Barut, 1984) is gauged, 
 the $\gamma $-matrices became function valued but remained
 covariantly constant with respect to the internal spin group, whereas the gravitational interaction is
 mediated by the four gauge-bosons belonging to the group $SU(2) \times U(1)$
 and the classical non-euclidian metric is constructed out of them as an effective field in a certain manner.

Here a modification in the sense of the Higgs-field gravity is
 indicated: Instead of considering the $\gamma $-matrices as
 covariantly constant it is possible to treat them as true field
 variables with a Higgs-Lagrange density, and this because also
 the $\gamma $-matrices possess a non-trivial ground-state, namely
 the usual constant standard representations. Because the 
$\gamma $-matrices can be understood as square root of the metric the
 gauge group is that of the square root of the metric; moreover,
 in consequence of this group the several spin states (or particle-antiparticle states) are indistinguishable with respect to the interaction following from gauging the spin group (universality of the interaction). Both properties suggest that real gravity is involved. 

In this way we get a quantizable unitary spin-gauge theory with
 Dirac's $\gamma $-matrices as Higgs-fields; on this level an
 unification with all the other interactions may be possible.
 After spontaneous symmetry breaking a non-scalar Higgs-gravity appears, which
 can be identified in a classical limit with Einstein's gravity,
 where we restrict ourselves in the first step for simplicity to the linear theory.
 The essential points are the following: the theory is from the beginning only Lorentz-covariant. After symmetry breaking and performing a unitary gauge the action of the excited $\gamma $-Higgs-field on the fermions in the Minkowski space-time is reinterpreted as if there would exist non-euclidean space-time connections and a non-euclidean metric (effective metric), in which the fermions move freely; then the deviation from the Minkowski space-time describes classical gravity. This happens, as usual, in the de Donder gauge and not in general coordinate covariance, which depends also on the fact that with the choice of the unitary gauge a gauge fixing is connected. In this way the  gravitational constant is produced only by the symmetry
 breaking and the non-euclidian metric comes out to be an effective field,  whereas
 the gauge-bosons get masses of the order of the Planck-mass and
 can be therefore neglected in the low energy limit; but in the
 high energy limit ($ \simeq 10 ^{19} $ GeV) an additional "strong" gravitational
 interaction exists. Simultaneously, our results give a new light
 on the role of the Higgs mechanism.

Finally we note, that as in the previous spin-gauge theory (c.f.
 Ghaboussi, Dehnen  and Israelit, 1987) a richer space-time
 geometrical structure results than only a Riemannian one. We find
 also an effective  non-metricity, whereas an effective torsion does not appear. The question, whether it is possible to
 change the Lagrangian so that  the non-metricity vanishes, will
 be clarified in a later paper.

\section*{II. The Model}
In the beginning we repeat briefly the foundations of the previous
 work (Dehnen, Ghaboussi, 1986;
 see also Babu Joseph, Sabir, 1988)
 so far as necessary. Using 4-spinors it is
 appropriate to introduce the transformation matrices of the group
 $SU(2) \times U(1)$ in their $ 4 \times 4$-representation $(a = 0,1,2,3):$
$$
U = e ^{i \lambda _a (x ^\mu ) \tau ^a} \, ,
\leqno (2.1)
$$
where the $SU(2)$-generators are given by the Pauli matrices
 $\sigma ^i $ as follows ($i = 1,2,3):$ \footnote{The explicite form of (2.2) is only used in (4.7).}
$$
\tau ^i = \frac{1}{2} \left( 
\begin{array}{cc}
\sigma ^i & 0 \\
0 & \sigma ^i 
\end{array}
\right) .
\leqno (2.2)
$$
The $U(1)$-generator $\tau ^0$ may be diagonal and commutes with (2.2); but its special form shall be determined only later. Thus the
 commutator relations for the generators $\tau ^a$ are
$$
 \left[ \tau ^b , \tau ^c \right] = i 
\epsilon _ a {}^{bc} \tau ^a , 
\leqno (2.3)
$$
where $\epsilon _ a {}^{bc} $ is the Levi-Civita symbol with the additional property to be zero, if $a$, $b$, or $c$ is  zero. 

Then the 4-spinor $\psi $ and the Dirac matrices $\gamma ^\mu $ transform as\footnote{$\gamma  ^\mu$ are tensors with respect to the unitary transformations (2.1), but they are not elements of the adjoint representation.}
$$
\psi ' = U \psi , \quad \gamma ^{' \mu } = 
U \gamma ^\mu U ^{-1} 
\leqno (2.4)
$$
and the covariant spinor derivative reads
$$
D_ \mu  \psi = (\partial _ \mu  + i g \omega _ \mu ) \psi 
\leqno (2.5)
$$
($g$ gauge coupling constant). The gauge potentials 
$\omega _ \mu $ obey the transformation law \footnote{$| \mu $ denotes the partial derivative with respect to the coordinate $x ^\mu $.}
$$
\omega ' _ \mu = U \omega _ \mu  U ^{-1} + \frac{i}{g} 
U _ {| \mu} U ^{-1} 
\leqno (2.6)
$$
and are connected with the real valued gauge fields 
$\omega _ {\mu a} $ by
$$
\omega _ \mu = \omega _ {\mu a} \tau ^a \, .
\leqno (2.7)
$$

According to (2.4) Dirac's $\gamma $-matrices become necessarly
 function valued, in consequence of which we need determination
 equations for them; as such ones we have chosen in our previous paper in analogy to general relativity:
$$
D_ \alpha \gamma ^\mu  = \partial _ \alpha  \gamma ^\mu  + 
ig \left[ \omega _ \alpha , \gamma ^\mu \right] = 0  \, , \quad 
\gamma ^{( \mu } \gamma ^{\nu )} = 
\eta ^{\mu \nu } \cdot \bf {1} \,  
\leqno (2.8)
$$
$( \eta ^{\mu \nu} \! = \! \eta _{\mu \nu}=\mbox{diag} (+1, -1, -1, -1)$ Minkowski metric). Because the $\gamma $-matrices are the formal square root of the
 metric, the gauge transformations (2.4) are those, which are
 associated with the root of the metric. Therefore the concept described by the formulae (2.1) up to (2.7) may have to do something with gravity. And indeed, in our previous paper we could show, that a space-time geometrical interpretation of the theory results in an effective  non-euclidian metric given by 
$$
g_ {\mu \nu } = \omega _ {\mu a} \omega _ {\nu b} \eta ^{a b} .
\leqno (2.9)
$$

However, the result (2.9) is connected with the condition that the
 gauge potentials $ \omega _ {\mu a}$ do never vanish and possess
 a non-trivial ground-state representing according to (2.9) in the
 lowest order the Minkowski metric. This is an  unusual feature;
 furthermore the conditions (2.8) are chosen for simplicity.
 Therefore it may be justified to give up the relations (2.8) and
 (2.9) and to consider Dirac's $\gamma $-matrices as true field
 variables with a Higgs-Lagrange density, so that the non-trivial
 ground-state can be identified with the constant standard representations.
 It will come out, that after symmetry breaking the excited
 $\gamma $-Higgs-fields mediate a non-scalar Higgs-gravity, which
 results
 finally in Einstein's metrical theory, where instead of (2.9)  the connection between the effective non-euclidian metric and the $\gamma $-Higgs-field will be deduced from a space-time geometrical interpretation
 of the equation of motion for the 4-momentum of the fermions
 described by the spinor fields $\psi $.

\section*{III. Lagrange Density and Field Equations }
The translation of the model into a field-theoretical description results in a
 Lagrange density consisting of three minimally coupled Lorentz- and gauge-invariant real valued
 parts ($ \hbar = 1,  \, c = 1)$:
$$
{\cal{L}} = {\cal{L}} _ M (\psi ) + {\cal{L}} _ F (\omega ) 
+ {\cal{L}} _H (\gamma ) \, .
\leqno (3.1)
$$
Beginning with the last part, ${\cal{L}} _H (\gamma )$ belongs to
 the $\gamma $-Higgs-field and has the form: 
$$
{\cal{L}} _H (\gamma ) =   \frac{1}{2} \mbox {tr}
 \left[ ( D_ \alpha 
 \, \tilde \gamma ^\mu ) (D^\alpha 
\, \tilde \gamma _ \mu ) \right] - 
V( \, \tilde \gamma ) - k \overline{\psi }\, \tilde \gamma ^\mu \, \tilde \gamma _\mu  \psi ,  
\leqno (3.2)
$$
where 
$$
V( \, \tilde \gamma ) = \frac{\mu ^2}{2}  \mbox {tr}
(\, \tilde \gamma ^\mu  \, \tilde \gamma _ \mu ) + 
\frac{\lambda }{4!} ( \mbox {tr} 
\, \tilde \gamma ^\mu  \, \tilde  \gamma _ \mu )^2 
\leqno (3.2a)
$$
is the Higgs-potential. 
Herein $\, \tilde \gamma ^\mu $ denotes from now the dynamic
 function valued $\gamma $-matrices, which obey the transformation law (2.4) and  the ground-states of which
 are proportional to the constant standard representations $\gamma ^\mu $ (bear this change of notation in mind). 
 The last term on the right hand side of (3.2) represents the
 Yukawa-coupling term for generating the mass of the fermions by the $\gamma $-Higgs-field. In view of the elektroweak interaction later on $\, \tilde \gamma ^\mu $ must become isospin valued, which leads to the possibility of unification in a 8-dimensional spin-isospin space (c.f. chapt. 6).

The second term on the right hand side of (3.1) is that of the
 gauge-fields $\omega _ \mu $:
$$
{\cal{L}} _ F (\omega ) = - \frac{1}{16 \pi } F_{\mu \nu a} 
F^ {\mu \nu } {}_b s^{ab} , 
\leqno (3.3)
$$
where $s^{ab}$ is the group-metric of $SU(2) \times U(1)$ and can
 be taken here as $\delta ^{ab}$ (but compare the previous work) .
 The gauge field strength are defined in the usual manner by
$$
{\cal{F}} _ {\mu \nu } = \frac{1}{ig} 
\left[ D_ \mu , D_ \nu \right] = F_ {\mu \nu a} \tau ^a 
\leqno (3.4)
$$
with 
$$
F_ {\mu \nu a} = \omega _ {\nu a | \mu } - \omega _ {\mu a | \nu } 
- g \epsilon _ a {}^{jk} \omega _ {\mu j} \omega _ {\nu k} \, .
\leqno (3.4a )
$$

The first Lagrangian in (3.1) concerns the fermionic matter fields
 and takes the form ($\psi $ is only proportional to the Dirac
 spinor, see (4.4): 
$$
{\cal{L}}_M (\psi ) = \frac{i}{2} \overline{\psi } \, \tilde \gamma ^\mu D_ \mu \psi - 
\frac{i}{2} ( \overline{D_ \mu \psi } ) \, 
\tilde \gamma ^\mu \psi . 
\leqno (3.5)
$$
The adjoint spinor $\overline{\psi }$ is given by 
$$
\overline{\psi } = \psi ^{\dagger} \zeta  , 
\leqno (3.6)
$$
wherein $\zeta $ represents the $SU(2) \times U(1)$-covariant matrix with the
 property:
$$
(\zeta  \, \tilde \gamma ^\mu ) ^{\dagger} = \zeta \, \tilde
 \gamma ^\mu \, .
\leqno (3.7)
$$
In view of the commutability of covariant derivative 
and multiplication with $\zeta  $ in (3.5) it is further
 necessary that 
$$
D_ \mu \zeta = 0 .
\leqno (3.8)
$$
So long as (see chapt. 4) 
$$
\left[ \gamma ^0 , \tau ^a \right] = 0 ,
\leqno (3.9)
$$
the equations (3.7) and (3.8)
will be fulfilled only (up to a constant factor) by 
$$
\zeta = \gamma ^0 \quad (\zeta ^ {\dagger} = \zeta , \, \zeta ^2 = 1) , 
\leqno (3.10)
$$
so that (3.6) yields as usual $\overline{\psi } = \psi ^{\dagger} \gamma ^0$. Because of (3.9) the matrix $\zeta $ is not only covariant but even invariant under gauge transformations. These results depend essentially on the relation (3.9),
 which may be not valid in a larger group (e.g. $U(4))$.
 \footnote{A generalization of the theory to the full gauge group U(4) is in
 preparation (see also Drechsler, 1988).} Finally we note, that one can prove easily with the use of (4.8), that
 all three expressions (3.2), (3.3) and (3.5) of the Lagrangian
 are real valued and contain \underline{no dimensional} parameter with
 exception of  $\mu ^2$ in (3.2), which has the dimension of a mass square. 

The field equations following from the action principle associated
 with (3.1) are given by the generalized Dirac-equation 
$$
i \, \tilde \gamma ^\mu  D_ \mu \psi  + \frac{i}{2} (D_ \mu \, \tilde \gamma ^\mu ) \psi  - 
k \, \tilde \gamma ^\mu \, \tilde \gamma _\mu  \psi = 0
\leqno (3.11)
$$
as well as its adjoint equation, by the inhomogeneous Yang-Mills
 equation
$$
\partial _ \nu F^ {\nu \mu } {}_a + g \epsilon _a {}^{bc} 
F^{\nu \mu }_ b \omega _ {\nu c} = 4 \pi j ^\mu  _ a
\leqno (3.12)
$$
with the gauge currents
$$
j ^\mu _ a = j ^\mu _a (\psi ) + j^\mu _ a (\gamma )
= \frac{g}{2} \overline{\psi } \left\{ \, \tilde \gamma  ^\mu  ,
 \tau _a \right\} \psi + 
$$
$$
+ ig \mbox {tr} \left( \left[ \, \tilde  \gamma ^\alpha , \tau _a \right]
D^\mu  \, \tilde  \gamma _\alpha  \right) 
\leqno (3.12a)
$$
belonging to the matter and the Higgs-field respectively, and by the $\gamma $-Higgs-field equation: \footnote{If $\, \tilde \gamma ^\mu  $ is considered to be traceless, see (4.2) and (4.8), then also the traceless version of (3.13) is valid only.}
$$
D_ \alpha D^\alpha \, \tilde \gamma ^\mu  {}_A {}^B + 
\left[ \mu ^2 + \frac{\lambda }{6} \mbox {tr} 
(\, \tilde \gamma ^\alpha \, \tilde \gamma _ \alpha ) 
\right] \, \tilde \gamma ^\mu _ A {}^B = 
$$
$$
= \frac{i}{2} \left[ \overline{\psi } ^B \cdot (D^\mu \psi )_ A -  (\overline{D^\mu \psi })^B \cdot \psi _A \right] - 
$$
$$
- k \left[\overline{\psi } ^B \cdot (\, \tilde \gamma ^\mu  \psi ) _ A
+ ( \overline{\psi } \, \tilde \gamma ^\mu ) ^B \cdot \psi _A \right] .  
\leqno (3.13)
$$
Herein the lower capital latin index A and the upper index B
 denote the contragradiently tranformed rows and collumns of the
 spinorial matrices respectively. The homogeneous Yang-Mills
 equation following from the Jacobi-identity reads: 
$$
\partial _ { \left[ \mu \right. } F_ {\left. \nu \lambda \right] a} 
+ g \omega _ {k \left[ \mu  \right. } F_ {\left. \nu \lambda  \right] j} 
 \epsilon ^{k j} {}_ {a} = 0. 
\leqno (3.14)
$$

Finally we note the conservation laws valid modulo the field equations. First, from (3.12) the gauge current conservation
 follows immediately: 
$$
\partial _ \mu  ( j ^\mu {}_a + 
\frac{g}{4 \pi } \epsilon _a {}^{bc} F^{\mu \nu } 
{}_ b \omega _ {\nu c} ) = 0 .
\leqno (3.15)
$$
Secondly, the energy-momentum law takes the form 
$$
\partial _ \nu T_ \mu {}^\nu  = 0 ,
\leqno (3.16)
$$
where $T_ \mu {}^\nu $ is the gauge-invariant canonical energy-momentum tensor consisting of three parts corresponding to (3.1)
$$
T_ \mu {}^\nu = T_ \mu {}^\nu (\psi ) + 
T_ \mu {}^\nu  ( \omega ) +
T_ \mu  {}^\nu ( \gamma )
\leqno (3.17)
$$
with (modulo Dirac-equation):
$$
T_ \mu  {}^\nu (\psi ) = 
\frac{i}{2} \left[ \overline{\psi } \, \tilde \gamma  ^\nu  
D_ \mu  \psi - 
(\overline{D_ \mu  \psi } ) \, \tilde  \gamma ^\nu  \psi  \right] , 
\leqno (3.18a)
$$
$$
T_ \mu  {}^\nu  ( \omega ) = - \frac{1}{4 \pi } \left[
F_ {\mu \alpha a} F^{\nu \alpha a} 
- \frac{1}{4} F_ a ^{\alpha \beta } F^a _ {\alpha \beta } 
\delta ^\nu _ \mu  \right], 
\leqno (3.18b)
$$
$$
T_ \mu  {}^\nu  (\gamma ) = \mbox{tr} 
\left[ (D^\nu  \, \tilde  \gamma  _ \alpha ) 
(D_ \mu  \, \tilde  \gamma  ^\alpha ) \right] - 
$$
$$
- \delta _ \mu  ^\nu  \left\{
\frac{1}{2} \mbox{tr} \left[ (D_ \alpha
 \, \tilde \gamma ^\beta )
( D^\alpha \, \tilde \gamma _ \beta ) \right] - 
\frac{\mu ^2}{2} \mbox{tr} ( \, \tilde  \gamma ^\alpha 
 \, \tilde  \gamma _ \alpha ) - \right.
$$
$$
- \left. \frac{\lambda }{4 !} ( \mbox{tr} \, \tilde \gamma ^\alpha 
\, \tilde \gamma _ \alpha ) ^2 \right\} .
\leqno (3.18c)
$$
Because of the Yukawa-coupling term in (3.2) the trace of (3.18a)
 does not vanish. With the use of the Dirac-equation (3.11) and
 its adjoint equation one finds: 
$$
T_ \mu  {}^\mu (\psi ) = k \overline{\psi }
  \, \tilde \gamma ^\mu \, \tilde \gamma _\mu \psi . 
\leqno (3.19)
$$

By insertion of (3.18) into (3.17) one obtains from (3.16) the
 equation of motion for the fermions. After substitution of the
 second covariant derivatives of the $\gamma $-Higgs-field using
 the field equation (3.13) one finds with the help of the 
Yang-Mills equations (3.12) and (3.14): 
$$
\begin{array}{ll}
\, & \partial _ \nu T ^{\mu \nu }(\psi ) = - 
\frac{i}{2} \left[ \overline{\psi } (D^\mu  
\, \tilde  \gamma ^\alpha ) ( D_ \alpha  \psi ) \right. - \\ [0.5 cm] 
\, & - (\overline{D_ \alpha  \psi } ) \left. ( D^\mu  \, \tilde \gamma ^\alpha ) 
\psi \right] + k \overline{\psi } \left\{  D^ \mu \, \tilde \gamma _\alpha  , \, \tilde \gamma ^\alpha \right\} \psi + \\ [0.5 cm]
\, &  + F^\mu  {}_ {\alpha  a} j^{\alpha  a} (\psi ). 
\end{array}
\leqno (3.20)
$$
Intergration over the space-like hypersurface $t$ = const. 
and neglection of surface integrals in the space-like infinity
 yield the momentum law for the 4-momentum $p ^\mu  = 
\int T^{\mu _0} (\psi ) d^3 x$ of the fermions. On the right hand 
side of (3.20) one recognizes the Lorentz-forces of the gauge fields 
and the force of the $\gamma $-Higgs-field. 

We finish with two remarks. First, the energy momentum tensor 
$T_ \mu  {}^\nu  (\gamma )$, equ. (3.18c), does not vanish for the ground-state, see (4.2), but has the value:
$$
\stackrel{(0)}{T} _ \mu {}^\nu ( \stackrel{(0)}{\gamma } ) = 
- \frac{3}{2} \frac{\mu ^4}{\lambda } \delta _ \mu  {}^\nu . 
\leqno (3.21)
$$
However this can be renormalized to zero by changing the Higgs-potential (3.2a) correspondingly; otherwise (3.21) will 
give rise within the complete theory to a cosmological 
constant. Secondly, the $\gamma $-Higgs-field equation 
(3.13) contains as source for $\, \tilde \gamma ^\mu $ 
the fermionic energy-momentum tensor 
$T_ \mu {}^\nu  (\psi )$ in its spinor valued form; and in this form it appears also in the $\gamma $-Higgs-field force of (3.20). 
This fact confirmes the supposition, that the 
$\gamma $-Higgs-field equation results in Einstein's 
field equation of gravitation (for the fermions) after  
a space-time geometrical interpretation of the 
$\gamma$-Higgs-field forces in (3.20) defining the 
effective space-time
 geometrical connection coefficients. 

\section*{IV. Spontaneous Symmetry Breaking}
Although one can recognize the gravitational structure already in equation (3.13) and (3.20) the space-time geometrical
 interpretation is only possible after symmetry breaking. 
The minimum of the energy-momentum tensor (3.18) in absence of matter and
 gauge fields is reached, when the Higgs-potential (3.2a) 
 is in its minimum defined by 
$$
\mbox{tr} \left( 
\stackrel{(0)}{\, \tilde \gamma ^\mu }  
\stackrel{(0)}{\, \tilde \gamma } _ \mu \right) = 
- \frac{6 \mu ^2 }{\lambda } = v^2 \quad 
(\mu ^2 < 0). 
\leqno (4.1)
$$
Simultaneously, herewith all field equations (3.11) up to (3.14) are fulfilled. The ground-state $ \stackrel{(0)}{\, \tilde \gamma ^\mu}  $ of the
 $\gamma$-Higgs-field must be proportional to the (constant)
 Dirac standard representation $\gamma ^\mu $, i.e. 
 $ \stackrel{(0)}{\, \tilde \gamma ^\mu } 
 = b \gamma ^\mu $.\footnote{
  Of course, global unitary transformations between the different
  standard representations and simultaneously of the generators
  are allowed.
 }
Insertion into (4.1) results because of $\left\{  \gamma ^\mu  , 
\gamma ^\nu \right\} 
= 2 \eta ^{\mu \nu } \cdot \bf{1} $ in $ b = \frac{v}{4}$, 
so that we have for the ground-state: 
$$
\stackrel{(0)}{\, \tilde \gamma ^\mu}  = \frac{v}{4} \gamma ^\mu . 
\leqno (4.2)
$$
Herewith the Lagrange density (3.5) for the spinorial matter fields 
reads considering the $\gamma $-Higgs-field ground-state only:
$$
\frac{i}{2} \overline{\psi } \frac{v}{4} \gamma  ^\mu \partial _ \mu  
\psi  + h. c. 
\leqno (4.3)
$$
Comparison with the usual Dirac Lagrangian 
$ \frac{i}{2} \overline{\psi }
_ {DIR} \gamma  ^\mu  \partial _ \mu  \psi _ {DIR} $ 
results in ($\psi _ {DIR}$  Dirac spinor): 
$$
\psi = \frac{2}{\sqrt v} \psi _ {DIR} . 
\leqno (4.4)
$$
Herewith the fermionic mass term in (3.2), identical 
with the trace (3.19) of the energy-momentum tensor 
$T_ \mu {}^\nu (\psi )$, takes the form for the groundstate $\stackrel{(0)}{\, \tilde \gamma}^\mu$ :
$$
T_ \mu  {}^\mu (\psi_ {DIR} )  = \overline{\psi } _ 
{DIR} m \psi _ {DIR} 
\leqno (4.5)
$$
with the mass:
$$
m = kv . 
\leqno (4.5a)
$$

On the other hand the Higgs-field gauge current 
$ j ^\mu  _ a (\gamma )$ gives rise after symmetry breaking to the mass 
of the gauge-bosons $ \omega  ^\mu  _a $. 
In the lowest order we find from (3.12a) with the use of (4.2):
$$
\begin{array}{ll}
\, & - 4 \pi  j ^\mu  _ a (\stackrel{0}{\gamma } ) = 
M^2 _ {ab} \omega  ^{\mu  b} , \; M^2 _ {ab} 
= M ^2 _ {ab} {}^{\rho \nu } \eta _ {\rho \nu } \, ,  \\ [0.5cm]
\, & M^2 _ {ab} {}^{\rho  \nu } = - 
\frac{\pi }{4} g^2 v^2 \mbox{tr}  
\left( \left[ \tau _a , \gamma  ^\rho \right]
 \left[ \tau _ b , \gamma  ^\nu \right] \right). 
\end{array}
\leqno (4.6)
$$

Here it is convenient to choose the $U(1)$-generator 
explicitely. If we take the unit matrix, the gauge-boson 
$\omega  ^\mu {}_ 0$ remains massless of course (rest symmetry) and must be taken into
 account also in the low energy limit. In order to avoid 
this, {\footnote{It seems to us not suitable to identify this 
boson with the photon in view of the electroweak interaction.}}
the first  possibility consists in view of (3.9) in the choice 
$\tau ^0 = \frac{1}{2} \gamma ^0$. Doing this we obtain from (4.6) with the use of (2.2)  the diagonal mass matrix for the gauge-bosons:
$$
 \begin{array}{ll}
\, & M ^2 _ {00}  =  3 \pi  g ^2 v ^2 , \\ [0.5 cm] 
\, & M ^2 _ {i j} =  2 \pi  g ^2 v ^2 \delta  _ {i j} 
\end{array}
\leqno (4.7)
$$
and zero otherwise. As we will see later the value of (4.7) 
is of the order of the square of the Planck-mass 
$( \stackrel{\wedge}{ = } 10 ^{19} $ GeV), so that all gauge-bosons can be neglected in the low energy limit. A second possibility for avoiding the $\omega ^\mu {}_ 0 $- boson exists in remaining the unit matrix for $\tau^0$ but choosing the associated gauge-coupling constant $g_1$ sufficiently small ($g_1 << 1$). This choice has the advantage, that the unitary $SU(2) \times U(1)$ transformation (2.1) up to (2.4) is exactly identical with that of the 2-spinors (resulting from a decomposition of  the chiral representation). 

As one can prove easily, the general Higgs-field $\, \tilde  \gamma  ^\mu $ can be represented, if no spin orientation is present (classical limit), by 
$$
\, \tilde  \gamma  ^\mu ( x ^\alpha ) = 
h^\mu  {}_\lambda (x^\nu ) U \stackrel{(0)}{\, \tilde \gamma  ^\lambda } U^{-1} , 
\leqno (4.8)
$$
so that 
it can be reduced within the unitary gauge as usual to the ground-state (4.2) in the following way
$$
\, \tilde  \gamma  ^\mu (x^\nu )  = h ^\mu {}_\lambda 
(x ^\nu ) 
\stackrel{(0)}{\, \tilde \gamma ^\lambda } ,
\leqno (4.8a)
$$
where \footnote{ Lifting and lowering of indices is performed
 always with $\eta  ^{\mu  \nu }$ and $ \eta _ {\nu \lambda  }$ respectively.} 
$$
h ^\mu  {}_ \lambda ( x ^\nu ) = \delta ^\mu  {}_ \lambda  + 
\epsilon ^\mu  {}_ \lambda (x ^\nu )
\leqno (4.8b)
$$
and $\epsilon ^\mu  {}_ \lambda  (x ^\nu )$ describes the
 deviations from the ground-state, i.e. the excited Higgs-field.
 Herewith we are able to write down all field equations after
 symmetry breaking exactly in a non-matrix valued form. Of course,
 the $h^\mu {}_\lambda (x^\nu )$ look like a tetrad-field, but
 their determination and connection with the effective 
non-euclidean metric follow only from the $\gamma$-Higgs-field
 equation after symmetry breaking. 

\section*{V. Field Equations after Symmetry breaking and Gravitational Interaction}
In this section we restrict ourselves in a first step for simplicity 
to the linearized theory, i.e. $ | \epsilon 
^\mu  {}_ \lambda | << 1$ (weak field limit). We start in view of the gravitational
 aspect with the Higgs-field equation (3.13). Going over from a
 spinorial description to a Lorentz-tensorial equation we multiply
 (3.13) at first by $\gamma ^\lambda  {}_ B {}^A$. Then after
 insertion of (4.2), (4.4), (4.5), (4.8a) and (4.8b) we obtain linearized
 in $\epsilon  ^\mu  {}_\lambda $ under neglection of the 
 gauge-boson interaction because of (4.7) (low energy limit): 
$$
\partial _\alpha  \partial ^\alpha \epsilon  ^{\mu \lambda } - \frac{\mu ^2 }{2} \epsilon  \eta  ^{\mu  \lambda } = \frac{4}{v^2} \left[
T ^{\mu \lambda } (\psi _ {DIR} ) - \frac{1}{2}
T(\psi _{DIR} ) \eta ^{\mu \lambda } \right] , 
\leqno (5.1)
$$
where
$$
T^{\mu \lambda } (\psi _ {DIR} ) = \frac{i}{2} \left[
\overline{\psi } _ {DIR} \gamma  ^\lambda  D^\mu \psi _ {DIR}
- ( \overline{D ^\mu  \psi} _ {DIR} ) \gamma ^\lambda \psi_ {DIR} \right]
\leqno (5.1a)
$$
is the usual (canonical) Dirac energy-momentum tensor. Obviously
 the antisymmetric and the traceless symmetric part of $\epsilon
 ^{\mu \lambda } $ remain massless, whereas the scalar trace $\epsilon = 
\epsilon ^ {\mu \lambda } \eta _ {\mu \lambda } $ possesses the Higgs-mass:
$$
M = \sqrt {- 2 \mu  ^2} . 
\leqno (5.1b)
$$
It seems, that in the standard model of electroweak interaction only this scalar part of the total Higgs-field is taken into account. Furthermore, if (5.1) shall describe usual gravity, $v^2 \sim G^{-1} $ ($G$ Newtonian gravitational constant) must be valid, 
so that (4.7) is indeed of the order of the square of the Planck-mass 
$M_ {Pl} = \frac{1}{\sqrt G}$. 

Before comparing (5.1) with Einstein's field equations it is 
appropriate to interpret at first the Higgs-field forces in (3.20) geometrically, 
where in the low energy limit the Lorentz-forces of the gauge fields can be neglected. 
Insertion of (4.2), (4.4), (4.8a) and (4.8b) into (3.20) gives with respect 
to (3.18a) and (5.1a): 
$$
\partial _ \nu  T^{\mu \nu } 
(\psi _ {DIR}) = - \partial  ^\nu  \epsilon _ {\nu \rho } 
T^{\mu \rho } 
(\psi _ {DIR} ) - 
\partial ^\mu  \epsilon _ {\rho \nu } 
T^{\rho \nu  } (\psi _ {DIR} ) + 
$$
$$
+ \frac{1}{2} \partial ^\mu \epsilon 
T(\psi _ {DIR}) 
\leqno (5.2)
$$
linearized with regard to $\epsilon ^\mu  {}_ \lambda $. 
The equations (5.1) and (5.2) describe the $\gamma $-Higgs-field 
interaction in its linearized version which is obviously very similar to that of general relativity.

Now, the comparison of (5.2) with the energy-momentum 
law of a classical affine geometrical theory with the 
affine connections $\Gamma^\mu  {}_{\nu \rho } $
$$
D_ \nu ^{ (\Gamma) } T^{\mu \nu } = 0 \Rightarrow \partial _ \nu T ^{\mu \nu } = 
-  \Gamma^\nu  {}_ {\nu \rho } T^{\mu \rho } -  \Gamma^\mu  {}_ {\nu  \rho } T^{\rho \nu } 
\leqno (5.3)
$$
is possible. Within this classical procedure we neglect all spin-influences, in consequence of which, c.f. (5.7), $T^{\rho \nu } (\psi _ {DIR} \equiv T^{\rho \nu } = T^{\nu \rho } $ is a symmetric tensor; thus we find the unique identification:
$$
\Gamma^\mu  {}_ {\nu \rho } = 
\partial ^\mu (\epsilon _ {\rho \nu } - \frac{1}{2} 
\epsilon \eta _  {\rho \nu } ) + 
\frac{1}{14} (\partial _\nu \epsilon \delta ^\mu _\rho + \partial _\rho \epsilon \delta _\nu {}^\mu )
\leqno (5.4)
$$
if
$$
\partial ^\nu  \epsilon _ {[\rho \nu ] } = 0 
\leqno (5.4a)
$$
is valid, see (5.6b). Herewith the forces of the excited $\gamma$-Higgs-field on the fermions in the Minkowski space-time are reinterpreted as the action of non-euclidean
 space-time geometrical connections. 

Consequently, in the space-time geometrical limit the excited 
Higgs-field $\epsilon ^\mu {}_\lambda $ or more precisely 
its derivatives play effectively the role of affine 
connections (effective connections). Their field equations 
are obtained in the following way: The equations 
(5.1) take the form assuming a negligible Higgs-mass (5.1b): 
$$
 \partial _ \alpha  \partial ^\alpha \left[ \epsilon _ {( \mu \nu) } 
- \frac{1}{2}\epsilon \eta _ {\mu \nu } \right] = 
\frac{4 }{v^2 } T_ {(\mu \nu )} ( \psi _ {DIR}) 
\leqno (5.5a)
$$
and 
$$
\partial _ \alpha \partial ^\alpha \epsilon _ {\left[ \mu \nu \right]} = 
\frac{4 }{v^2} T_ {\left[ \mu \nu \right]  } ( \psi _ {DIR}) . 
\leqno (5.5b)
$$
In the lowest order, which is considered here only, the right 
hand sides of (5.5) possess in view of (5.2) vanishing divergences.
 Therefore the following constraints hold in consequence of the
 field equations: 
$$
\begin{array}{ll}
\mbox{(a)} & \hspace{0,5cm}
 \partial ^\nu  \left[ \epsilon _ {(\mu  \nu )}  - \frac{1}{2}
\epsilon \eta _{\mu \nu } \right]  =   0 , \\
\mbox{(b)} & \hspace{0,5cm}
\partial ^\nu \epsilon _ {\left[ \mu \nu \right] }  =  0,  \, 
\end{array}
\leqno (5.6)
$$
the first of which has the structure of the de Donder condition and the second of which guarantees the fulfillment of the condition (5.4a).

Evidently the source of the antisymmetric part $\epsilon _ {[\mu \nu ]}$  is the antisymmetric part of
 the fermionic energy-momentum tensor (5.1a), which can be written
 with the use of the Dirac-equation in the lowest order: 
$$
T_ {\left[\mu \nu \right]} ( \psi _ {DIR} ) = 
\frac{1}{2} 
\left[ \overline{\psi}  _ {DIR} \gamma  _ {\left[\mu  \right.}
 \sigma  _ {\nu  \left. \right]} 
{}^\lambda  D_ \lambda \psi _ {DIR} \right. + 
$$
$$
+ \left.  (\overline{D_ \lambda \psi } _ {DIR} ) \sigma  ^\lambda 
 {}_ {\left[ \mu  \right. } \gamma  _ {\nu \left. \right]} 
\psi _ {DIR} \right] , 
\leqno (5.7)
$$
where $\sigma ^{ \mu \nu  } = i \gamma  ^{\left[ \mu \right. } \gamma^{\left. \nu  \right]} $ is the spin-operator.  Therefore, if we neglect in the classical macroscopic
 limit all spin influences, the solution of (5.5b) is: 
$$
\epsilon _ {\left[ \mu \nu \right]} \equiv 0, 
\leqno (5.8)
$$
whereby in the classical limit $\epsilon _ {\mu \nu } = 
\epsilon _ {(\mu \nu )}$ is valid. 

For discussion of the field equation (5.5a) for the symmetric part
 $\epsilon _ {(\mu \nu )}$  we compare directly with 
 Einstein's linearized field equations of gravity. Setting
$$
g_ {\mu \nu } = \eta  _ {\mu \nu } + \gamma _ {\mu \nu } 
\leqno (5.9)
$$
and choosing the de Donder gauge, c.f. (5.6a)
$$
\partial ^\nu  ( \gamma _ {\mu \nu } 
- {\textstyle{\frac{1}{2}}}  \gamma  \eta _ {\mu \nu } ) = 0 
\leqno (5.9a)
$$
($ \gamma = \gamma  _ {\mu \nu } \eta ^{\mu \nu }$) it is valid:
$$
\partial _ \alpha  \partial ^\alpha  {\textstyle{\frac{1}{2}}} ( \gamma _ {\mu \nu } - 
{\textstyle{\frac{1}{2}}} \gamma \eta _ {\mu \nu } ) = - 8 \pi G T_ {( \mu \nu )} \, .
\leqno (5.10)
$$

The comparison with (5.5a) results immediately in:
$$
\epsilon _ {(\mu \nu )} = \alpha \gamma _ {\mu \nu } 
\leqno (5.11)
$$
and 
$$
v^2 = (- 4 \pi G \alpha )^{-1}  
\leqno (5.11a)
$$
up to the proportional constant $\alpha $. Consequently the constraint (5.6a) is identical with the
 de Donder condition (5.9a) and Newton's gravitational constant $G$ is correlated, as expected,  with the Higgs-field ground-state value $v$. The constant $\alpha $ is adjusted in such a way, that the equation of motion (5.2) goes over in the lowest order into the Newtonian gravitational law; for this 
$$
\epsilon _ {(\mu \nu )}  = - \frac{1}{4} \gamma _ {\mu \nu }  \Rightarrow \alpha = - \frac{1}{4} , \quad v^2 = 1/\pi G 
\leqno (5.12)
$$
must be valid in view of (5.11) with $\gamma _ {\mu \nu } = 
2 \Phi{\mbox{diag}} (1,1,1,1) $ with respect to  (5.10) ($\Phi$ Newtonian gravitational potential). 
Herewith
 the effective non-euclidian metric takes the form with respect to
 (5.9): 
$$
 g_ {\mu \nu } = \eta _ {\mu \nu } - 4 \epsilon _ {(\mu \nu )}. 
\leqno (5.13)
$$

Analyzing the affine geometric connections (5.4) we note that beside the Christoffel-symbols belonging to the metric (5.13) 
$$
\left\{  
\alpha \atop {\mu \nu } 
\right\} 
= 2 \eta ^{\alpha \lambda } 
( \partial _\lambda \epsilon _ {(\mu \nu) } - 
\partial _\nu \epsilon _ {(\mu \lambda )}
- \partial _\mu  \epsilon _ {(\lambda \nu )}) 
\leqno (5.14)
$$
there exists no effective torsion
$$
 \Gamma^\mu {}_ {[\nu \rho ]} \equiv 0
\leqno (5.15)
$$
but effective non-metricity:
$$
Q_ {\mu \nu \lambda } = 
- D_ \mu ^{( \Gamma)} g_ {\nu \lambda} = 
4 \partial _ \mu \epsilon _ {(\nu \lambda )} + 
\partial _\lambda \epsilon _{(\nu \mu )} + \partial _\nu \epsilon _{(\lambda \mu )} - 
$$
$$
 -  \frac{3}{7}(\partial _\nu \epsilon \eta_ {\mu \lambda } + 
\partial _\lambda \epsilon \eta _{\mu \nu } ) + 
\frac{1}{7} \partial _\mu \epsilon \eta _{\nu \lambda }
\leqno (5.16)
$$
assuming $\epsilon _{[\mu \nu ]} \equiv  0$ in all cases, c.f. (5.8). 

The practical consequences of the appearance of  non-metricity or even better its avoidance shall be investigated elsewhere.

Now  we note the Dirac-equation for gravitational interaction
 according to the spin-gauge theory as well as the Yang-Mills
 equation for the very massive gauge fields. From (3.11) it
 follows immediately after insertion of (4.2), (4.4), (4.5a) and
 (4.8) under neglection of the gauge-boson interaction
 (low energy limit):
$$
i \gamma ^\mu {\cal{D}} _\mu \psi _ {DIR} - 
m (1 + \frac{1}{2}\epsilon ) \psi _ {DIR} = 0 
\leqno (5.17)
$$
with 
$$
{\cal{D}}_\mu = \partial _\mu + \epsilon ^\lambda {}_\mu \partial _\lambda + \frac{1}{2} 
(\partial _\lambda \epsilon ^\lambda {}_\mu ) . 
\leqno (5.17a)
$$
In its non-relativistic limit this equation goes over into the Schr\"odinger-equation with usual Newtonian gravitational potential. Iteration of (5.17), elimination of all spin influences and linearization in $\epsilon ^\lambda {}_ \mu $ give: {\footnote{We use $\hbar , c$ explicitely because of the ordering with respect to $c^{-1}$.}} 
$$
{\cal{D}}^\mu {\cal{D}}_ \mu \psi _ {DIR} + 
\frac{m^2 c^2}{\hbar^2} ( 1 + \epsilon ) \psi _ {DIR} + 
\frac{i}{2} \frac{mc}{\hbar} \gamma ^\mu 
(\partial _\mu \epsilon ) \psi _ {DIR} = 0.
\leqno (5.18)
$$
With the ansatz
$$
\psi _ {DIR} = \epsilon ^ {- i \frac{mc^2}{\hbar}t} 
\varphi (x^\nu )
\leqno (5.19)
$$
we obtain from (5.18) under neglection of all terms up to the order of $c^{-1} (\epsilon ^\lambda {}_\mu   \sim c^{-2})$ the Schr\"odinger-equation 
$$
\frac{\hbar^2}{2m} \Delta \varphi + 
\frac{mc^2}{2} (2 \epsilon ^{00} - 
\epsilon ) \varphi = 
\frac{\hbar}{i} \partial _t \varphi 
\leqno (5.20)
$$
Because of (5.12) $\epsilon ^{00} = - \frac{1}{2}\Phi /c^2 $, $\epsilon = \Phi /c^2$ is valid, so that (5.20) goes over into 
$$
\frac{\hbar^2}{2 m} \Delta \varphi - 
m \Phi \varphi = 
\frac{\hbar}{i} \partial _t \varphi  , 
\leqno (5.21)
$$
i.e. the usual Schr\"odinger-equation with classical gravitational potential $\Phi$.

We have shown this explicitely, because this quantum
 mechanical equation has been tested experimentally until now only
 for the gravitational interaction by the neutron-interference experiment of Collela, Overhauser and Werner (1975). 
It may be of interest however, that the Schr\"odinger-equation  (5.21) 
does not guarantee, that atomic clocks and lengths measure the
 effective non-Euclidean metric; for this the influence of the
 gravitational field on the electric Coulomb potential 
between electron and nucleus of the atom is necessary 
(see e.g. Papapetrou, 1956), which is not yet included in our theory.

Finally, for the inhomogeneous Yang-Mills equation we obtain from (3.12) 
and (3.12a) with the use of (4.2), (4.4), (4.6) and (4.8):
$$
\partial _ \nu  F^{\nu \mu } {}_a +
g \epsilon _a {}^{bc} F^ {\nu \mu } {}_ b \omega _ {\nu c} +
M^2 _ {ab} \omega ^{\mu b} + 
2 \epsilon _ {(\rho \nu )} M^2 _ {ab}{}^{\rho \nu } \omega ^{\mu b} = 
$$
$$
= 4 \pi  \left\{  \frac{g}{2} \overline{\psi }_ {DIR} 
\left\{  \gamma ^\lambda  , \tau  _ a \right\} 
\psi _ {DIR} ( \delta   ^\mu {}_ \lambda + \epsilon  ^\mu  {}_ \lambda ) + \right. 
$$
$$
\left.  i g \frac{v^2}{16} ( \partial ^\mu  \epsilon  _ {\rho \nu }) 
\mbox {tr} \left( \left[ \gamma ^\rho  , \tau  _ a \right] \gamma  ^\nu  \right) \right\} , 
\leqno (5.22)
$$
where we have restricted ourselves also to linearized interaction
 terms with respect to gravitation. On the left hand side we recognize the
 mass term and the interaction of the massive bosons with the
 gravitational potentials; on the right hand side we find as sources 
 gravitationally influenced Dirac gauge currents and a current
 associated with the gravitational field itself (remaining Higgs-field
 current). Because of this it may be justified to call the gauge-boson interaction as a "strong" but very massive gravitational
 interaction; its coupling constant $g$ remains however
 undetermined within our present theoretical approach.

\section*{VI. Final Remarks}
In extension of a previous spin-gauge theory of gravity we have shown, that Dirac's $\gamma $-matrices can be treated as a
 quantizable Higgs-field, in consequence of which Einstein's
 metrical theory of gravitation follows as the classical
 macroscopic limit of the Higgs-field interaction after symmetry
 breaking. 

In spite of this success there are several problems for the
 future. First, the effective space-time geometrical structure 
is not only a Riemannian one, but also non-metricity is present,
 which should be suppressed in the next step since no
 observational hint on it exists. This may be possible because the
 Lagrange density (3.2) for the Higgs-field is not yet unique but
 can be supplemented in its kinetic term, e.g. by 
tr$[(D^\alpha  \, \tilde \gamma ^\mu )(D_ \mu
 \, \tilde  \gamma ^\alpha )]$. In connection with this it may 
also be attainable to avoid the constraint (5.6a), which
 corresponds to the de Donder condition, and perhaps in this 
way Einstein's theory can be reached even exactly and not only
 in its linearized version as presented above. 

Furthermore the theory, as it stands, contains only the
 gravitational interaction between the fermions. But the
 gravitational interaction with all bosons must be included 
within a complete and consistent theory of gravitation; otherwise, as remarked above (c. f. Papapetrou, (1956)), atomic clocks and
 lengths do not measure the non-Euclidean effective metric. 
This may require however an unification with the other
 interactions on the microscopic level of unitary phase 
gauge transformations within a high dimensional (e.g. 
8-dimensional) spin-isospin space describing gravitational 
and (electro-) weak interaction separately. 

In this respect one could have a bold idea: Because in our 
theory the $\gamma $-matrices are treated as Higgs-field it 
could be possible to introduce the chiral asymmetry of the
 fermions with regard to the weak interaction, which is 
however present already in the $SU(5)-GUT$, by a special 
choice of the ground-state of the $\gamma $-Higgs-field in 
the course of the spontaneous symmetry breaking at 
approximately $10^{19}$ GeV connected with the 
separation of gravitational and electro-weak interaction. 

\section*{References}

\begin{itemize}
\item[] Bade, W. , Jehle, H., Rev. Mod. Phys. \underline{25}, 714 (1953); see also \\
Laporte, O., Uhlenbeck, B., Phys. Rev. \underline{37}, 1380 (1931) 
\item[] Babu Joseph, K., Sabir, M., Mod. Phys. Lett. A
 \underline{3}, 497 (1988)
\item[] Barut, A.O., McEwan, J.,
 Phys. Lett \underline{135}, 172 (1984);
 Lett. Math. Phys. \underline{11}, 67 (1986)
\item[] Chisholm, J., Farwell, R., J. Phys. \underline{A22}, 1059 (1989), and the literature cited therein
\item[] Collela, R., Overhauser, A., Werner, S., Phys. Rev. Lett. \underline{34}, 1472 (1975)
\item[] Dehnen, H., Ghaboussi, F., Nucl. Phys. \underline{B262}, 144 (1985)
\item[] Dehnen, H., Ghaboussi, F., Phys. Rev. \underline{D33}, 2205 (1986)
\item[] Dehnen, H., Frommert, H., Ghaboussi, F., Int. J. theor. Phys. \underline{29}, 537 (1990)
\item[] Dehnen, H., Frommert, H., Int. J. theor. Phys. \underline{30}, 985 (1991)
\item[] Drechsler, W., Z. Phys. \underline{C41}, 197 (1988)
\item[] Ghaboussi, F., Dehnen, H., Israelit, M., Phys. Rev. \underline{D35}, 1189 (1987)
\item[] Ghaboussi, F., Il Nuov. Cim. \underline{104A}, 1475 (1991)
\item[] Papapetrou, A., Ann. der Phys. \underline{17}, 214, (1956)
\item[] Schouten, J., Ricci-Calculus, 2nd edition, Springer (Berlin, 1954)
\item[] Stumpf, H., Z. Naturforsch. \underline{43a}, 345 (1988)
\end{itemize}

\end{document}